\documentstyle[preprint,prb,aps]{revtex}
\begin{document}
\draft
\title{Dynamical correlations in a Hubbard chain with
       resonating-valence-bond ground state}
\author{Andreas Giesekus}
\address{Institut f\"ur Physik, Universit\"at Dortmund\\
         D-44221 Dortmund\\
         Germany}
\maketitle
\begin{abstract}
Dynamical correlation functions for temperature $T=0$ are
calculated for a Hubbard chain with infinite on-site repulsion.
This chain contains three sites per unit cell and has a known
resonating-valence-bond ground state for a filling of 2 particles
per unit cell. A finite system of 24 sites is studied numerically.
The recursion method is applied and results for spectral densities
are compared to variationally calculated excited-state eigenvalues.
Charge and spin degrees of freedom located at the backbone sites
are considered. The spectral densities show dispersionless
excitations. The energy spectrum  displays a gap indicating
an insulating ground state. Spectral functions for the propagation
of a single hole in a resonating-valence-bond background are
provided and indicate delocalized dispersive hole excitations.
\end{abstract}

\narrowtext
%

%
\section{INTRODUCTION}
\label{sec:1}

The Hubbard model\cite{hubbard} describes electronic
interactions in tight binding systems. Although the
model itself is very simple, the access to its
mathematical properties turns out to be very difficult.
The present work is based on an exactly solvable
case of the Hubbard model in the strongly interacting
limit ($U=\infty$). For that case it has been possible
to construct the ground state and the ground-state
energy.\cite{brandt-giesekus}
The class of models which are solvable by this method
has been generalized by Strack,\cite{strack} Bares and
Lee\cite{bares-lee} and Tasaki.\cite{tasaki1,tasaki2}

The ground state of those models exhibits the structure
of a resonating-valence-bond (RVB) state.\cite{tasaki1}
One-dimensional models of this class\cite{strack} allowed
for the calculation of equal-time correlation
functions\cite{bares-lee} using a transfer-matrix technique.

All studies conducted so far have been concerned with equilibrium
properties of the ground state. Although the explicit knowledge
of a Hamiltonian for the fermionic realization of an RVB-system
enables one to study the excitation spectrum and dynamical
quantities, no analytical statements about those quantities have
been possible up to now.

This paper presents a numerical treatment of excitations in a
one-dimensional representative of the class of exactly solvable
models introduced in [\onlinecite{brandt-giesekus}].
Dynamical correlation functions are calculated by a recursive
algorithm. The resulting spectral properties are corroborated by
a variational calculation of excited states and their energies.
The method used here allows the calculation of single-hole
spectral-functions as well. In principle these spectral functions
are related to inverse photoemission spectra.

\section{The model Hamiltonian}
\label{sec:2}

A linear chain with $N$ unit cells and $3N$ sites is
studied. The topology of this chain is illustrated in
Fig.\ \ref{fig:1}; it may be interpreted as a chain of
connected tetrahedra.
The case of infinite on-site repulsion is
assumed, which implies that only particle numbers
below half-filling exhibit non-trivial dynamics. The
ground state can be constructed for particle densities
of two or more per unit cell,\cite{brandt-giesekus}
thus the chain must have at least three sites (labeled by
indices $\alpha$, $\beta$) per unit cell.

The dimensionless Hamiltonian (the hopping matrix element
defines the energy scale only and is set to unity)
of the system under investigation is written
as:
\begin{eqnarray}
\label{eq:1}
{\rm H} \quad = \quad {\rm P} \, \Big\{
              &-& \,\,  \sum_{i,\sigma}
                    \sum_{\alpha \not= \beta}
                             {c_{i,\alpha,\sigma}^{\dagger}}
                              c_{i,\beta,\sigma}^{}\\
\nonumber
             &-& \,\, \sum_{i,\sigma}\sum_{\alpha}\big(
                             {c_{i-1,1,\sigma}^{\dagger}}
                              c_{i,\alpha,\sigma}^{} + h.c. \big)\\
             &-& \,\, 2 \, \sum_{i,\sigma} n_{i,1,\sigma}^{}
                                  \Big\} \,{\rm P},
\nonumber
\end{eqnarray}
where ${c_{i,\alpha,\sigma}^{\dagger}}$ creates an
electron in the unit cell $i\in \{1,\ldots,N\}$ at site
$\alpha\in\{1,2,3\}$ with spin
$\sigma\in\{\uparrow ,\downarrow\}$. The operators
$n_{i,\alpha,\sigma}$ denote the corresponding occupation
number operators. Periodic boundary conditions are assumed.
As mentioned above, an infinitely strong on-site interaction
is included by projecting all states onto the subspace
without doubly occupied sites:
\begin{equation}
\label{eq:2}
{\rm P} \quad = \quad \prod_{i,\alpha}\big(
            1 - n_{i,\alpha,\uparrow}^{}n_{i,\alpha,\downarrow}^{}
                                       \big)
\end{equation}

The last term in (\ref{eq:1}) is a local field that
lowers the energy of the connection sites $\alpha=1$.
It ensures an almost uniform distribution of particles
in the chain: The ground-state probability for finding an
electron with spin $\sigma$ on a site with $\alpha=1$
($\alpha=2,3$) in a particular unit cell has been
calculated as 0.3612 (0.3194) for a chain with six unit
cells.

\section{The ground state}
\label{sec:3}

The ground state is derived by the same arguments as
presented in [\onlinecite{brandt-giesekus}].
Consider a linear combination of the creation operators
$c_{i,\alpha,\sigma}^{\dagger}$:
\begin{equation}
\label{eq:3}
{\Psi}_{i,\sigma}^{\dagger} :=
c_{i-1,1,\sigma}^{\dagger}
\,+\,
\sum_{\alpha} c_{i,\alpha,\sigma}^{\dagger}
\end{equation}
By some elementary algebra, the Hamiltonian (\ref{eq:1})
can be exactly converted to the following form
\begin{equation}
\label{eq:4}
{\rm H} = {\rm P} \, \left\{
                         2 {{\rm N}}_e - 8N
                   \right\}
             +
                \sum_{i,\sigma} {\Psi}_{i,\sigma}^{} {\rm P}
                               {\Psi}_{i,\sigma}^{\dagger}\,,
\end{equation}
where ${\rm N}_e$ denotes the total electron number
operator of the system. The Hamiltonian now consists
of a trivial term ${\rm P}(2{\rm N}_e - 8N)$ and a
remainder. This remainder is positive semidefinite,
therefore the ground-state energy is bounded from below
by $E_0\ge2N_e-8N$, where $N_e$ denotes the eigenvalue
of the electron-number operator ${\rm N}_e$. Additionally,
an upper bound can be derived: Any expectation value of
the Hamiltonian (\ref{eq:4}) is greater than or equal to
the ground-state energy. The operator identity
\begin{equation}
\label{eq:5}
{\rm P}
{\Psi}_{i,\sigma}^{\dagger}
{\rm P}
{\Psi}_{i,\sigma}^{\dagger}
\quad \equiv \quad 0
\end{equation}
allows to construct states which are eigenstates of the
remainder operator
$
\sum_{i,\sigma} {\Psi}_{i,\sigma}^{}
                {\rm P}
                {\Psi}_{i,\sigma}^{\dagger}
$
with eigenvalue zero. Any state
\begin{equation}
\label{eq:6}
|\Phi_0\rangle := {\rm P}
          \prod_i {\Psi}_{i,\uparrow}^{\dagger}
                  {\Psi}_{i,\downarrow}^{\dagger} |\chi\rangle
\end{equation}
has this property. If $|\chi\rangle$ denotes the
vacuum state $|0\rangle$, $|\Phi_0\rangle$ is simply
a Gutzwiller-projected Slater determinant with two
electronic orbitals in each unit cell.

The solvability of the model is based on Eq.\ (\ref{eq:5}),
a relation that holds because a local summation over
fermionic operators vanishes due to their phase and the
topology of the unit cell. This local property will be
visible in the spectral results of the ground-state
dynamics studied below.

It is mentioned in passing that the norm of the states
(\ref{eq:6}) is not zero, a fact that is not trivial and
has been proven.\cite{brandt-giesekus} For any finite
system, the ground state with $|\chi\rangle = |0\rangle$
is non-degenerate. This has been conjectured by Brandt
and Giesekus\cite{brandt-giesekus} and proven by
Tasaki\cite{tasaki2} (and Bares and Lee\cite{bares-lee}
for a special case), whereas a ground state with more than
$2N$ particles is obviously degenerate.

Tasaki\cite{tasaki1} pointed out that the ground state is
generated by linear combinations of creation operators that
contain only terms like
$
c_{i,\alpha,\uparrow}^{\dagger}
c_{j,\beta,\downarrow}^{\dagger} -
c_{i,\alpha,\downarrow}^{\dagger}
c_{j,\beta,\uparrow}^{\dagger}
$.
These terms create spin singlets on bonds connecting sites
$i,\alpha$ and $j,\beta$. Linear combinations with this
structure are often called (short range) resonating
valence-bond (RVB) states.\cite{pauling}

\section{Symmetries}
\label{sec:4}

For a numerical study of the linear chain it is essential
to take advantage of symmetries.
They allow to represent the states in terms of a significantly
smaller basis set, because only one configuration need be stored
as a representative of all states that may be generated from
it by symmetry operations.

\subsection{Site interchange}

The Hamiltonian is invariant under interchange of sites
$\alpha=2$ and $\alpha=3$ (see Fig.\,1) in every unit cell
$i$. As a consequence there exists an abelian group of $2^N$
symmetry operations that commute with the Hamilton operator.
All $2^N$ combinations of site interchanges are generated by
every single of the $2^N$ following operators:
\begin{equation}
\label{eq:7}
{\rm\bf M}(\{\epsilon_i\}) = \frac{1}{2^N}
\prod_{i=1}^N ({\rm I} + \epsilon_i {\rm M}_i)
\end{equation}
where ${\rm M}_i$ denotes site interchange in unit cell $i$
and ${\rm I}$ the unit operator. (The quantity $\epsilon_i$ can take
the values $\pm1$. Note that the product of two such operators vanishes
unless they coincide in all signs $\epsilon_i$.)

It is completely equivalent to interpret the {\em interchange
of sites} in terms of an {\em interchange of particles}, namely
those particles that are located on the two participating sites.
Then the properties of ${\rm M}_i$ acting on fermionic states
become obvious immediately: Let the states
$|00\rangle$,
$|\sigma0\rangle$,
$|0\sigma\rangle$,
$|\sigma\sigma'\rangle$
denote empty, singly and doubly occupied bonds.
The operator ${\rm M}_i$ has the following properties:
\begin{itemize}
\item ${\rm M}_i |00\rangle = |00\rangle$
\item ${\rm M}_i |\sigma0\rangle = |0\sigma\rangle$,
      ${\rm M}_i |0\sigma\rangle = |\sigma0\rangle$
\item ${\rm M}_i |\sigma\sigma'\rangle = -|\sigma'\sigma\rangle$.
\end{itemize}
{}From these properties, it follows immediately that the operator
$\frac{1}{2}({\rm I} + {\rm M}_i)$ not only generates the
site interchange, it additionally may be interpreted as a projector
onto spin singlets for doubly occupied bonds. The operator
$\frac{1}{2}({\rm I} - {\rm M}_i)$ projects onto triplet states.

The symmetry of every eigenstate of the Hamiltonian may be classified
according to $2^N$ different sets of signs $\{\epsilon_i\}$
[see Eq.\ (\ref{eq:7})].
The ground state may be written as
\begin{equation}
\label{eq:8}
|\Phi_0\rangle = {\rm\bf M}_s |\tilde{\Phi}_0\rangle \,.
\end{equation}
where ${\rm\bf M}_s$ (the index $s$ indicates that all $2-3$-bonds
are projected on the singlet-subspace) is an abbreviation for
${\rm\bf M}(\{\epsilon_i=+1\})$.
This is possible, because the ground state consists of singlets only.
However, it is important to keep in mind that excited states may not
necessarily be written in this way. The state
$|\tilde{\Phi}_0\rangle$ requires much less computer memory,
because it is only necessary to store one representative basis
state for the whole class of about $2^N$ basis states that are
connected by a simple interchange of sites $\alpha=2,3$.

\subsection{Translational invariance}

The second kind of symmetries that is taken explicitely
into account is the translational invariance of the Hamiltonian.
Periodic boundary conditions are assumed. Let ${\rm T}_l$
be the operator that shifts all electrons on the chain $l$
unit cells to the right. Again, fermion states are totally
antisymmetric under interchange of particles, hence translation
of states (wich may be described in terms of particle interchange)
may cause a phase. For the case of $l=1$ (shift by one unit cell)
the resulting sign is given by $(-1)^{n(N_e-n)}$, where $N_e$
is the total number of electrons in the system and $n$ the
number of electrons in the last unit cell that are transferred
to the first unit cell (i.\,e.\ moved across the ``border'').
The operator
\begin{equation}
\label{eq:9}
{\rm\bf T}_q = \frac{1}{\sqrt{N}}\sum_{l=1}^N {\rm e}^{iql} T_l
\end{equation}
acts as a generator of all translations with symmetry
$q=2\pi m/N$ ($m=0,\ldots,N-1$). It commutes
with the Hamiltonian and with ${\rm\bf M}_s$. (This is not
valid for {\em all} site-interchange symmetries
${\rm\bf M}(\{\epsilon_i\})$. The operator ${\rm\bf T}_q$
obeys the relation
${\rm\bf T}_q {\rm\bf T}_{q'} = \delta_{q,q'}{\rm\bf T}_q$.

The ground state $|\Phi_0\rangle$ may now be coded by
\begin{equation}
\label{eq:10}
|\Phi_0\rangle = {\rm\bf T}_{q=0}\,{\rm\bf M}_s
     |\bar{\Phi}_0\rangle \,,
\end{equation}
where $|\bar{\Phi}_0\rangle$ contains only those basis
states that may not be generated from each other by a
translation.

It is mentioned in passing that these symmetries lead
to selection rules: The scalar product of states can be
easily treated if one state has the site-interchange
symmetry ${\rm\bf M}_s$: The scalar product of the
states
$|f\rangle={\rm\bf T}_q {\rm\bf M}_s |\bar{f}\rangle$
and
$|g\rangle={\rm\bf T}_{q'} {\rm\bf M}(\{\epsilon_i\}) |\bar{g}\rangle$
does not vanish if $q=q'$ {\em and} all $\epsilon_i=1$
because ${\rm\bf M}_s$ and ${\rm\bf T}_q$ commute.
This special case is sufficient for the algorithms
used below. More general selection rules would have
to be derived by group theory because the total group
of symmetry operations (site-interchange and translations)
is not abelian.

The net storage reduction factor is of the order
of $N2^N$ (in fact, the number is slightly smaller,
because some basis components may be eigenstates
of ${\rm M}_i$ [e.\,g.\ empty bonds] or ${\rm T}_l$)
which reduces the multiplicity of the representative.
With these considerations, system sizes of 24 Hubbard
sites (eight unit cells) can be treated easily on a
workstation.

\section{Dynamical correlation functions}
\label{sec:5}

This section briefly recalls how dynamical correlation
functions at zero temperature are accessible by the
recursion method.\cite{haydock,lee,gagliano,viswanath_jap,pettifor}
A detailed discussion can be found in a very recent book by
Viswanath and M\"uller.\cite{vs-gm} Correlation
functions of the following type are considered:
\begin{equation}
\label{eq:11}
\tilde{S}(t) = \frac{\langle\Phi_0|{\rm A}^{\dagger}(t)
                                   {\rm A}(0)|\Phi_0\rangle}
                    {\langle\Phi_0|{\rm A}^{\dagger}(0)
                                   {\rm A}(0)|\Phi_0\rangle}\,.
\end{equation}
The operator ${\rm A}$ decribes the degree of freedom
under consideration (${\rm A}(t)$ denotes the Heisenberg picture).
Below, data are presented for three different ${\rm A}$-operators,
e.\,g.\ ${\rm A}={\rm S}^{z}_{q,\alpha=1}$, which measures
the $z$-component of the spin located at the connection
sites $\alpha=1$ with a translational symmetry $q$. Charge
degrees of freedom are studied by e.\,g.\
$
{\rm A}=\sum_{l,\sigma}
\exp\{ {\rm i} ql\}\,n_{l,\alpha,\sigma}
$.
Because the recursion method is not restricted to hermitean
${\rm A}$-operators, the propagation of a single hole may be
studied by the operator
$
{\rm A}=\sum_l
\exp\{ {\rm i} ql\}\,c_{l,\alpha,\sigma}
$. Then $\tilde{S}(t)$ [Eq.\ (\ref{eq:11})] becomes a single-particle
Green function (for $t>0$).

All ${\rm A}$-operators which are considered here commute with the
symmetry generator ${\rm\bf M}_s$ of Eq.\ (\ref{eq:8}) because they
only act on backbone orbitals (degrees of freedom regarding the
remaining sites would have to be symmetrized such that they commute
with ${\rm\bf M}_s$ as well). Therefore the following calculation
can be  performed without loss of generality in the subspace
corresponding  to this symmetry. This can be seen immediately
by formally expanding the state $\langle \Phi_0|{\rm A}^{\dagger}$
in terms of eigenstates $\langle\Phi_{\nu}|$ with coefficients
$\alpha_{\nu}$. Then the numerator of Eq.\ (\ref{eq:11}) becomes
$
\sum_{\nu}\,\alpha_{\nu} \exp (-{\rm i}E_{\nu}t/{\hbar})
\langle\Phi_{\nu}|{\rm A}|\Phi_0 \rangle
$
where $E_{\nu}$ denotes the eigenvalue. The matrix element vanishes
if the symmetries of ${\rm A}|\Phi_0 \rangle$ and $|\Phi_{\nu}\rangle$
differ.

The method described below requires a positive semidefinite
Hamiltonian. Therfore a shifted Hamiltonian
${\rm \bar{H}}={\rm H}-E_0$ is considered where the smallest
eigenvalue equals zero. For the hole propagator, the Hamiltonian
should be positive semidefinite for particle numbers $N_e=2N-1$.
Unfortunately the  ground-state energy is not known for that case,
however, the operator ${\rm \bar{H}}={\rm H}+4N+2$ is larger than
$\lambda\ge0$, where $\lambda$ is the smallest eigenvalue of the
non-trivial remainder in (\ref{eq:4}) for $N_e=2N-1$.

Dynamical correlation functions may be derived by an expansion
of the state
$
|\Psi (t)\rangle = {\rm A}(-t)|\Phi_0\rangle
$
which solves the time dependent Schr\"odinger equation.
This state $|\Psi (t)\rangle$ may be expanded in terms
of an orthogonal basis\cite{gagliano}
$|\Psi (t)\rangle=\sum_k D_k(t)|f_k\rangle$.
Then the time evolution of the coefficients $D_k(t)$
is governed by a set of coupled differential equations
which may be solved by a Laplace transform.

If this orthogonal basis $|f_k\rangle$ is obtained by
the following algorithm (which is closely related to
the Lanczos method\cite{lanczos})
\begin{mathletters}
\label{eq:12}
\begin{eqnarray}
|f_0\rangle&=&{\rm A}|\Phi_0\rangle\\
|f_1\rangle&=&{\rm \bar{H}}|f_0\rangle - a_0|f_0\rangle\,,\\
|f_2\rangle&=&{\rm \bar{H}}|f_1\rangle - a_1|f_1\rangle
                                 - b_1^2|f_0\rangle\,,\\
\nonumber
             &\vdots&\\
|f_{k+1}\rangle&=&{\rm \bar{H}}|f_k\rangle - a_k|f_k\rangle
                                     - b_k^2|f_{k-1}\rangle\, ,
\end{eqnarray}
\end{mathletters}
where
\begin{equation}
\label{eq:13}
a_k=\frac{\langle f_k|{\rm \bar{H}}|f_k\rangle}
             {\langle f_k|f_k\rangle}
\quad \text{and} \quad
b_k^2=\frac{\langle f_k|f_k\rangle}
       {\langle f_{k-1}|f_{k-1}\rangle}\, ,
\end{equation}
the coefficient $D_0 (t)$ is identical to $\tilde{S}(t)$
and the series of coefficients $a_k$, $b_k^2$ become
continued fraction coefficients of the Laplace transform
of $\tilde{S}(t)$:
\begin{eqnarray}
\nonumber
d_0 (\xi) &:=& \int_0^{\infty} {\rm d}t\,{\rm e}^{{\rm i}\xi t}
\tilde{S}(t)\\
\label{eq:14}
          & =&
\frac{\displaystyle {\rm i}}
     {\displaystyle \xi - a_0 - \frac{\displaystyle b_1^2}
     {\displaystyle \xi - a_1 - \frac{\displaystyle b_2^2}
     {\displaystyle \xi - a_2 \cdots}
}}\,.
\end{eqnarray}
If all coefficients $a_k$, $b_k^2$ are known, the
spectral density
$
S(\omega)=\int_{-\infty}^{\infty} {\rm d}t
\exp({\rm i}\omega t)\,\tilde{S}(t)
$
corresponding to the dynamical correlation function
(\ref{eq:11}) can be recovered by
\begin{equation}
\label{eq:15}
S(\omega)= \lim_{\delta \rightarrow 0}
2\,\Re[d_0(\omega+{\rm i}\delta)]\,,
\end{equation}
however, only a finite number of continued fraction
coefficients can be calculated. Nevertheless, the
first few coefficients contain valuable information about
the dynamics of the system which will be extracted (as
will be discussed in Sec.\ \ref{sec:6}) by more
sophisticated means then just cutting off the continued
fraction and smearing out by a finite imaginary smear-out
parameter $\delta$. A cut-off procedure simply approximates
a branch cut by a finite number of isolated singularities,
where the number of peaks is determined by the
order of the approximation. Additionally, the smear-out
parameter $\delta$ is often of the same order of magnitude
as the hopping matrix element, otherwise reasonably
looking results cannot be obtained. In fact, $\delta$ must be
larger than a typical distance between the isolated
singularities in order to produce a sufficient smear-out effect.

Finite Fermi systems have a discrete energy spectrum which
only becomes continuous in the thermodynamic limit. Thus a
recursive study of a finite Fermi system results in principle
in a finite continued fraction: The recursion will terminate,
if the states $|f_k\rangle$ span the Hilbert space completely.
In practice, this termination is only seen for very small
systems, because the number of eigenstates grows exponentially
with the size of the system.

\subsection{Remarks on the numerical implementation}

The sequence of coefficients $a_k$, $b_k^2$ have to be computed
numerically. In order to give a rough estimate of the numerical
effort required one should consider the maximum dimension $D$ of the
Hilbert space: The system considered contains $3N$ sites. $N$ particles
with spin up may be distributed on these sites in $({3N \atop N})$
possible ways, the remaining $N$ spin-down particles may be located
at the $2N$ unoccupied sites. Then the total dimension will
be $(3N)!/(N!)^3$ which amounts to $\approx 10^{10}$ for $N=8$.

In a first step, the ground state is constructed in terms of a basis
with binary coding of the orbital occupation. For the implementation
of the recursion it is essential to generate the ground state with
maximum precision. Here, the ground state is known, e.\,g.\, we know
an exact algorithm which generates the ground state by successive
creation of particles out of the vacuum, according to
Eq.\ (\ref{eq:6}). The coefficients of the ground-state
basis elements are coded as integers. This eliminates all
rounding errors.

The main part of the recursion Eqs. (\ref{eq:12}) is the
implementation of the Hamilton operator. Whenever ${\rm \bar{H}}$
acts on a component of a state $|f\rangle$,
${\rm \bar{H}}|f\rangle = |f'\rangle$, it is necessary to check
whether the resulting basis components have already been
generated before, in which case only amplitudes have to
be added. Otherwise the newly generated basis components
of $|f'\rangle$ have to be stored along with their amplitudes.
This simple operation requires a large share of the numerical
effort. In our algorithm this task is accelerated by application
of binary trees. Then the overall numerical effort is of order
$D\, \ln D'$, where $D$ is the dimension of the input state
$|f\rangle$ and $D'$ the dimension of $|f'\rangle$.
As stated above, $D$ and $D'$ increase exponentially with system
size. By taking advantage of the symmetries mentioned above,
a chain length of eight unit cells can be treated on a
workstation with 128 Mbyte of computer memory. In that case
$D$ is reduced to an amount of $\approx 10^6$ if symmetries
are taken into account. Without explicit use of symmetries,
$N=6$ is the maximum system size.

\section{Finite size effects and \protect\\
reconstruction of spectral densities}
\label{sec:6}

This section provides a brief recall how the information
contained in the first continued-fraction coefficients
can be reliably retrieved. Two problems need to be addressed:
First, because of limited memory and CPU-power only finite
systems can be studied, and, second, only a finite number
of continued-fraction coefficients can be calculated.

The method applied here has been developed in the framework
of an equivalent formulation (called Liouvillian
representation) of the recursive algorithm outlined in
Eqs.\,(\ref{eq:12}). This representation is based on an
orthogonal expansion\cite{lee} of the operator ${\rm A}(t)$
and leads to a continued fraction of the form
\begin{eqnarray}
\nonumber
c_0 (z) &:=& \int_0^{\infty}{\rm d} \,t \,\,
              {\rm e}^{-zt}\tilde{S}(t)\\
\label{eq:16}
&=& \frac{\displaystyle 1}
         {\displaystyle z+\frac{\displaystyle \Delta_1}
         {\displaystyle z+\frac{\displaystyle \Delta_2}
         {\displaystyle z+ \cdots}
}}\,.
\end{eqnarray}
The functions $S(\omega)$ and $c_0(z)$ are connected by a
Hilbert transform:
\begin{eqnarray}
\label{eq:17}
c_0 (z) &=& \frac{1}{2\pi {\rm i}}
            \int_{-\infty}^{\infty} {\rm d}\omega\,\,
            \frac{S(\omega)}{\omega-{\rm i}z}\\
\label{eq:18}
S(\omega) &=&  \lim_{\epsilon \to 0}
               2\,\Re[c_0(\epsilon - {\rm i}\omega)]
\end{eqnarray}
The series of $\Delta_k$ can be uniquely obtained from the
$a_k$-$b_k^2$-series.

\subsection{Finite size effects}

Finite size effects can be eliminated by investigation of
various data sets of $\Delta$-coefficients with varying
system size. The upper graph of Fig.\ \ref{fig:2} shows some
data sets for chain lengths from four up to eight unit cells.
The physical quantity under consideration is the spin correlation
function with ${\rm A}={\rm S}^{z}_{q,\alpha=1}$. The data
clearly illustrate that the finite size effects grow with
increasing index $k$ of $\Delta_k$. Therefore it is essential
to restrict the analysis to coefficients that are in good
agreement with those corresponding to shorter systems and
ignore the size dependent higher coefficients.

\subsection{Termination}

The full spectral information $S(\omega)$ is only accessible
if all coefficients $\Delta_k$ (of an infinite system) are
known. The question of how much information is contained in
a finite $\Delta_k$-series is by no means trivial and the
goal of constructing a good approximation to the infinite
continued fraction requires some care.

Let $\Gamma_K(z)$ be a terminator function that approximates
the unknown rest of the continued fraction in ``the best
possible way''.
\begin{equation}
\label{eq:21}
\bar{c}_0 (z) = \frac{\displaystyle 1}
         {\displaystyle z+\frac{\displaystyle \Delta_1}
         {\displaystyle z+\frac{\displaystyle \Delta_2}
         {\displaystyle z + \cdots
                \raisebox{-4ex}{$
                \cdots + \frac{\displaystyle \Delta_{K-1}}
                    {\displaystyle z + \Delta_K \Gamma_K (z)}
                                $}
         }}}
\end{equation}

The crudest approximation, $\Gamma_K (z) \equiv 0$, is
equivalent to the ``cut and smear out'' procedure mentioned
above. More adequate terminator functions may be obtained
by analyzing the {\em implicit} information contained in the
asymptotic behavior of the $\Delta_k$-series.

In order to construct a more adequate terminator function,
the pattern of $\Delta$-coefficients for many generic situations
have been analyzed in [\onlinecite{vs-gm}]; the corresponding
terminator functions  employed here are also discussed in
[\onlinecite{viswanath}]; the reader is referred to these sources
for more technical information.

Generally, linear growth of $\Delta_k$ with $k$ signals a
Gaussian high-frequency decay of $S(\omega)$. Linear growth
of both $\Delta_{2k}$ and $\Delta_{2k+1}$ with different
slopes occurs if a spectral gap at low frequencies is
present in $S(\omega)$. An example of this is shown in the
lower panel of Fig.\ \ref{fig:2} [line a)]. Line b) in the
same panel illustrates the characteristic change in the
roles of even and odd $\Delta$'s when an additional
zero-frequency peak is present in $S(\omega)$.

The size-independent numerically-cal\-cu\-la\-ted
con\-tinued-fraction coefficients of the system under
investigation show the generic patterns decribed above.
Thus a model spectral function which corresponds to these
patterns,
$
\bar{S}(\omega)=(2\sqrt{\pi}/\omega_0)
                \Theta (\omega-\omega_g)
                \exp\{-(\omega-\omega_g)^2/\omega_0^2\}
$,
is expanded into a model continued fraction. The gap-parameter
$\omega_g$ and the width of the shifted half-Gaussian $\omega_0$
are chosen such that the coefficients of the model continued
fraction match the coefficients of the real system as close as
possible. The terminator function $\Gamma_K (z)$ may be recursively
obtained from the Hilbert transform of $\bar{S}(\omega)$.\cite{hilbert}

The spectral density is reconstructed using all size-independent
coefficients calculated numerically and -- instead of just
truncating -- extrapolating the asymptotic pattern to infinite
order by ``implanting'' the customized terminator function. The
more size independent explicitly known coefficients are taken,
the less important are the details of the terminator. The main
advantage of this termination procedure is the fact that a
branch cut (in this formulation located on the imaginary axis
of the complex plane) of $c_0(z)$ is {\em not} approximated by
a finite set of isolated singularities; instead the function
$\bar{c}_0 (z)$ displays a branch cut itself.

The reconstruction of spectral densities is numerically
performed by Eq.\ (\ref{eq:18}). The parameter $\epsilon$
is chosen as $\epsilon=10^{-3}$. Reconstruction of the spectral
densities with $\epsilon=10^{-4}$ or $\epsilon=10^{-5}$ yields
results which coincide within the linewidth of
Figs.\ \ref{fig:3}-\ref{fig:4}.

\section{Excited states}
\label{sec:7}

The spectral densities obtained for spin- and charge degrees
of freedom indicate that the excitation spectrum of the
chain with $2N$ particles exhibits a gap or an energy region,
where the spectral densities have very little
weight. In order to gain complementary information
on the excitation spectrum of the system under consideration,
excited states and their corresponding eigenvalues have been
calculated by an iterative algorithm often called conjugate-gradient
method.\cite{bradbury,nightingale}
If the starting vector overlaps with the ground state, the
algorithm will iteratively determine the ground state. The same
procedure will yield excited states and excitation energies, if it
is carried out in a subspace orthogonal to the ground state
(provided the ground state is non-degenerate). Additionally,
excited states are iterativley calculated in different symmetry
subspaces. Fig.\ \ref{fig:5} provides the energy difference
between the ground state and the lowest excited states as a function
of system size.

\section{Results and Discussion}
\label{sec:8}

The present work investigates a Hubbard model of which the
ground state and its energy can be calculated for a special
number of particles $N_e=2N$. This ground state has an
RVB-structure. As far as the author knows, no other class of
fermionic Hamiltonians with such a state beeing the exact
ground state is known so far.

However, not many of the physical properties of this ground
state and the excitations are known. Static correlation
functions [the denominator in Eq.\ (\ref{eq:11})] have been
calculated for other one-dimensional representatives of this
class\cite{bares-lee} by a transfer-matrix method. In all cases
considered so far, the static correlation functions show an
exponential decay.

The method presented above allows to calculate wave-vector
dependent dynamical correlation functions making explicit
use of the fact that the ground state is known. If the energy
spectrum of the Hamiltonian has a gap, the spectral densities
will show a gap as well. (The gap might be larger than the gap
of the energy spectrum. In that case, this larger gap will be
called {\em dynamical} gap.) The position of peaks in the
reconstructed spectral densities as shown below reflect the
dispersion relation $\omega_{peak}(q)$ of that particular
excitation.

In order to corroborate the conclusions about a gap in the
excitation spectrum, energies of the lowest excited states
(without symmetry  restriction) and those excited states
that couple to the ground state are calculated.

Additionally, an attempt is  made to gain information on the
particle number regime below $N_e=2N$ which has not been studied
so far.

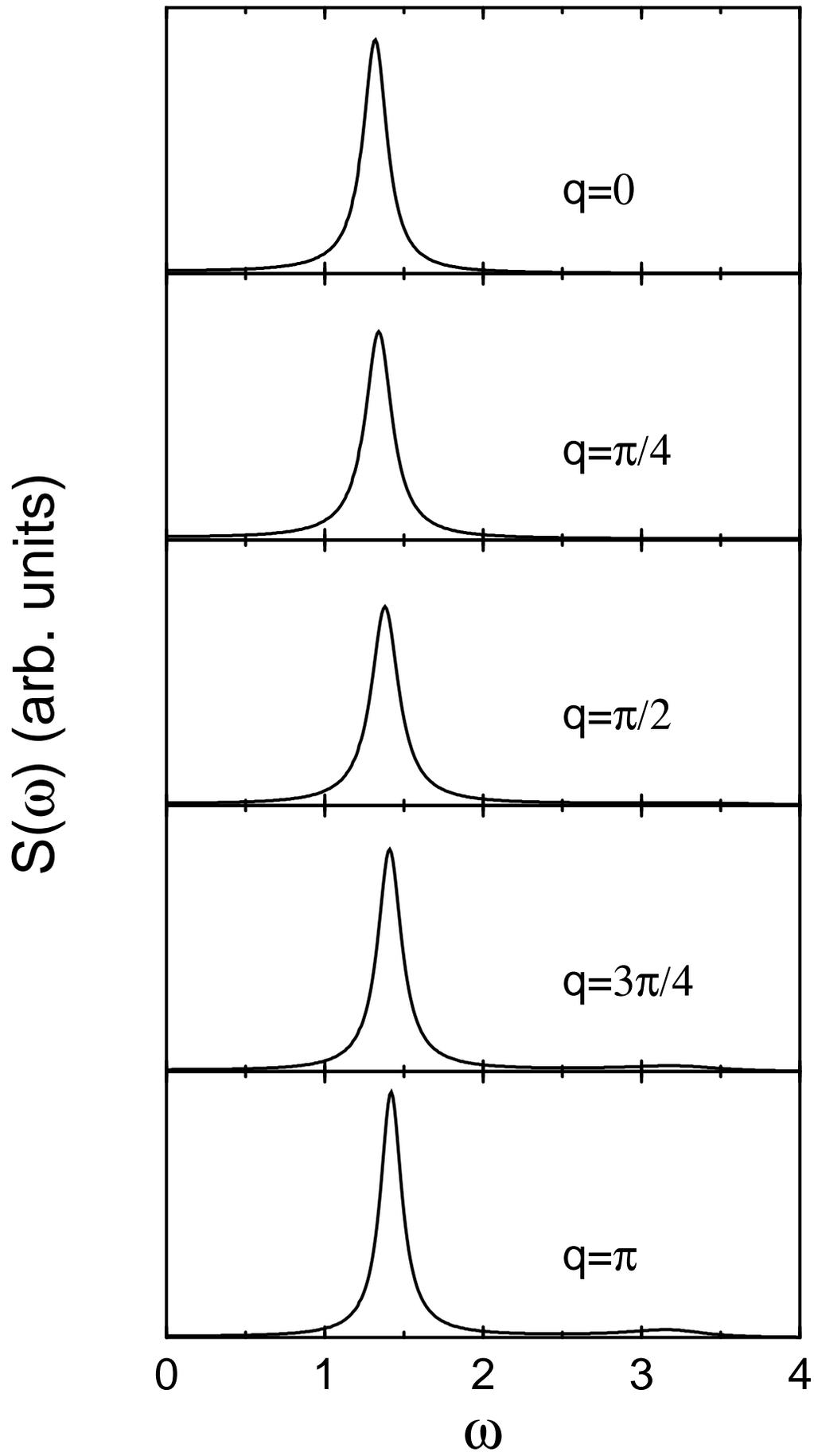
\begin{figure}
\caption{Reconstructed spectral densities (Gaussian terminator,
$\epsilon=10^{-3}$) for
spin degrees of freedom
${\rm A}_q=\sum_l \exp\{{\rm i} ql\}\,S^{z}_{l,\alpha=1}$.
The first 16 continued-fraction coefficients calculated for
the chain with $N=8$ were taken into account.
}
\label{fig:3}
\end{figure}

Fig.\ \ref{fig:3} shows spectral densities
$
S(\omega)=\int_{-\infty}^{\infty}{\rm d}t\,\exp({\rm i}\omega t)
\tilde{S}(t)
$ [see Eq.\ (\ref{eq:11})] for a spin degree of freedom
$
{\rm A} = \sum_l \exp({\rm i}ql) (n_{l,\alpha=1,\uparrow}
-n_{l,\alpha=1,\downarrow})
$
which measures the $z$-component of the spin of the backbone
electrons. All energies $\omega$ are given in units of the
hopping matrix element ($\hbar$ is set to unity). The translational
symmetry is characterized by $q=2\pi m/N$ with $m=0 \ldots N/2$.
The data indicate that there is a (dynamical) gap in the spectrum.
There is one dominating spin excitation present with q-independent
energy (dispersionless). Both facts support the conjecture, that
the system has an insulating\cite{bares-lee} ground state
and that the properties are dominated by local effects.
Only a very small spectral feature appears for $q\to\pi$
in the energy regime $\omega\approx 3$. All curves are
normalized to $\pi$.

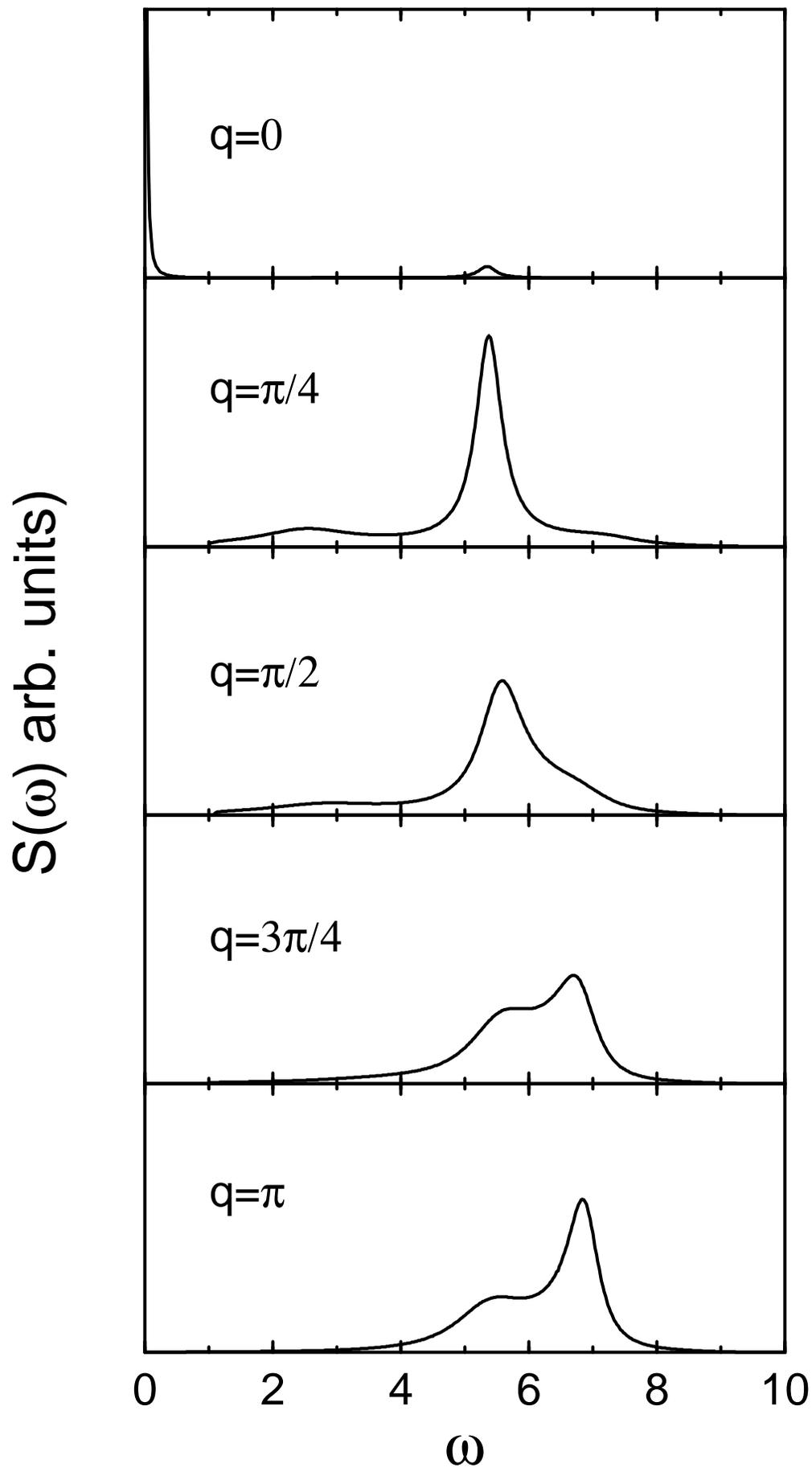
\begin{figure}
\caption{
Reconstructed spectral densities  (Gaussian terminator
[with gap for $q>0$], $\epsilon=10^{-3}$)
for charge degrees of freedom
${\rm A}_q=\sum_{l,\sigma} \exp\{{\rm i} ql\}\,n^{}_{l,\alpha=1,\sigma}$.
The first 16 continued-fraction coefficients calculated for
the chain with $N=8$ were taken into account.
}
\label{fig:4}
\end{figure}

The operator
${\rm A}$ $=$ $\sum_{l,\sigma} \exp({\rm i}ql) n_{l,\alpha=1,\sigma}$
describes charge degrees of freedom of the backbone sites.
The corresponding spectral density exhibits more than just
one dominating excitation. For $q=0$, the data in
Fig.\ \ref{fig:4} show one very strong peak at $\omega=0$.
This simply reflects the $q=0$-symmetry of the ground state.
For other $q$-values, features at higher energies appear.
Their position does not change much with $q$, the corresponding
excitations seem to be rather dispersionless.

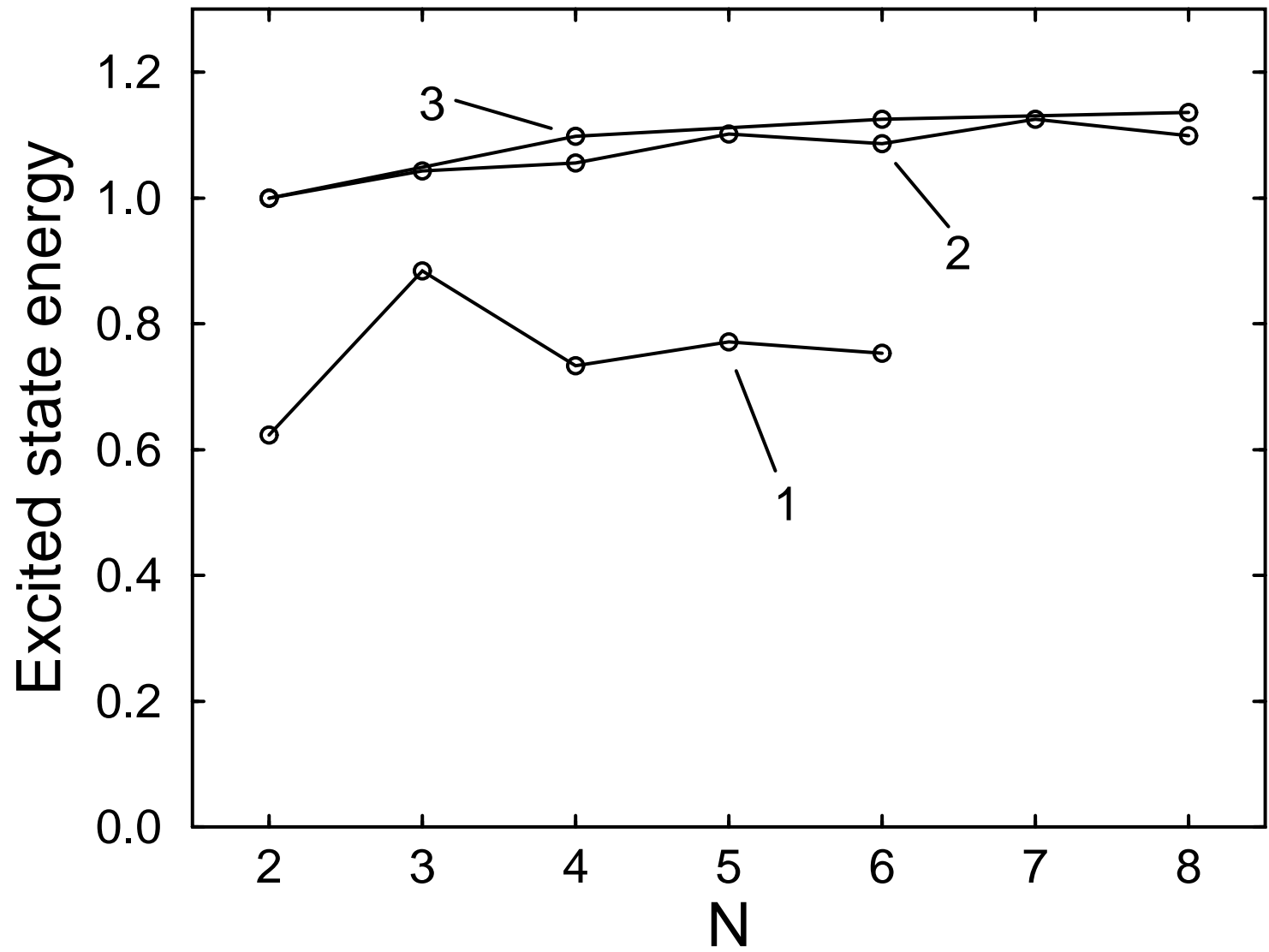
\begin{figure}
\caption{
Energy differences $E - E_0$ between excited-state
eigenvalues and the ground-state energy for systems with $N$
unit cells and $2N$ particles. Data set 1 provides results
without any symmetry restrictions, results in set 2 (set 3)
are based on the assumption that the corresponding excited
state $|\Psi\rangle$ has the symmetry
${\rm\bf T}_{q=0} {\rm\bf M}_s|\bar{\Psi}\rangle$
(${\rm\bf T}_{q=\pi} {\rm\bf M}_s|\bar{\Psi}\rangle$). Lines are
intended to guide the eye.}
\label{fig:5}
\end{figure}

The {\em dynamically relevant} gap is determined by the energy
of the lowest excited state which couples via ${\rm A}$ to the
ground state. Because the ${\rm A}$-operators considered here
commute with the singlet generator ${\rm\bf M}_s$
[see Eq.\ (\ref{eq:8})], the lowest accessible excited state must
have the symmetry ${\rm\bf T}_q {\rm\bf M}_s$ which is the symmetry
of the state ${\rm A}(q)|\Phi_0\rangle$. The data sets 2 and 3 in
Fig.\ \ref{fig:5} show these excitation energies as a function
of the system size. They seem to saturate at a constant level
which indicates the existence of a dynamical gap. This conclusion
is fully compatible to the reconstructed spectral data.
However, the energies of the lowest excited states {\em without any
restrictions on their symmetry} are located within the dynamical gap.
Data set 1 provides these energies for chain lengths up to $N=6$.
A final judgement, whether these energies saturate or not
cannot be made, because larger systems cannot be treated without
taking advantage of symmetries.

Further, the variationally calculated energies show only very
little dependency on the wave vector $q$ which confirms the very
small shift of the peak in Fig.\ \ref{fig:3}.

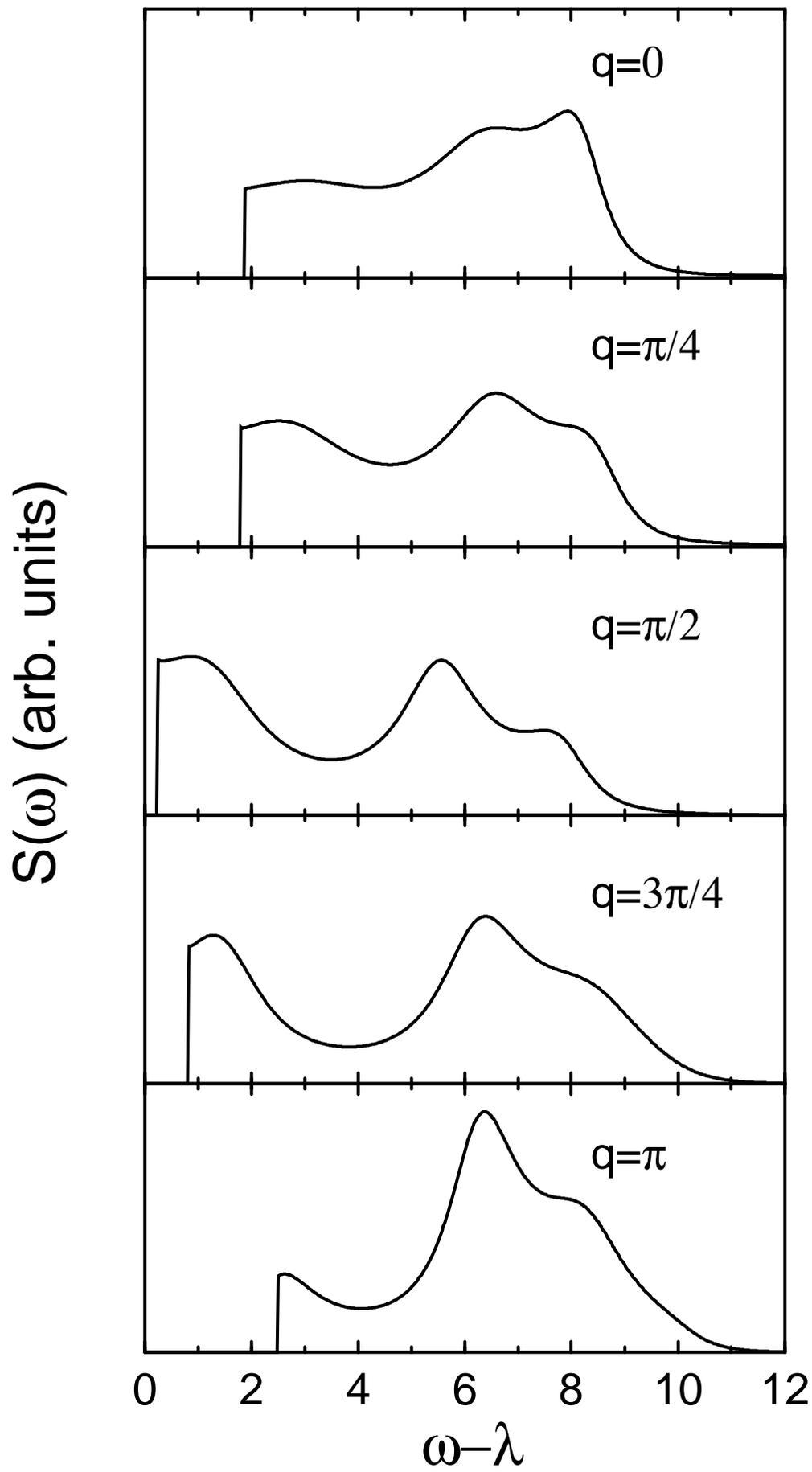
\begin{figure}
\caption{
Reconstructed spectral densities  (Gaussian terminator with gap,
$\epsilon=10^{-3}$) for the hole propagator
${\rm A}_q=\sum_l \exp\{{\rm i} ql\}\,c^{}_{l,\alpha=1,\sigma}$.
The first 16 continued-fraction coefficients calculated for
the chain with $N=8$ were taken into account. The origin of the
$\omega$-axis is shifted by the unknown smallest eigenvalue $\lambda$
of ${\rm \bar{H}}$ for $N_e=2N-1$
}
\label{fig:6}
\end{figure}

The recursion method is not restricted to hermitean operators
${\rm A}$ only. Fig.\ \ref{fig:6} provides results for
${\rm A}=\sum_l \exp({\rm i}ql)c_{l,\alpha=1,\sigma}^{}$ which
annihilates backbone electrons in a translationally invariant way.
Then the quantity $\tilde{S}(t)$ of Eq.\ (\ref{eq:11}) becomes a
single-hole propagator. Its Fourier transform may be related
to inverse photoemission spectra.

Characteristic features of the single-hole spectral function
shown in Fig.\ \ref{eq:6} are a broad continuum of states and a
$q$-dependent gap. Unfortunately the  origin of the $\omega$-axis
is not known, because the smallest  eigenvalue $\lambda$ of the
Hamiltonian ${\rm \bar{H}}={\rm H}+4N+2$ cannot be calculated. The
single-hole states are delocalized (i.\,e.\ show dispersion) and
may contribute to electronic conduction.

\acknowledgments

The author is indebted to J.\,Stolze who generously shared
his experience and knowledge about continued fractions and
their termination. His ``toolbox'' allowed to reconstruct
the spectral densities efficiently and has been of valuable
help. J.\,Richter was so kind to provide some of his computer
power during software development. The author gratefully
acknowledges many helpful discussions with W.\,Wenzel,
H.\,Keiter and U.\,Brandt. This work has been supported by
the Deutsche Forschungsgemeinschaft (DFG), Project Br\,434/6-2.

%
%

%
%

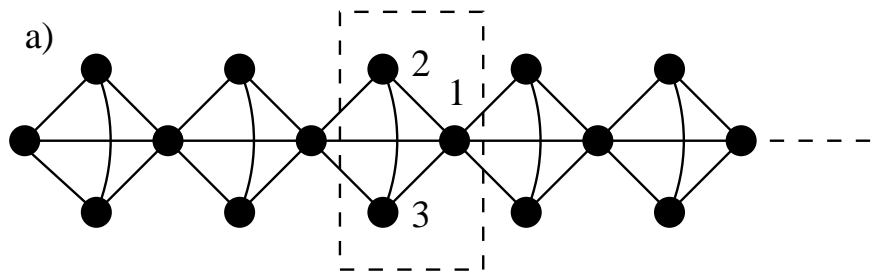
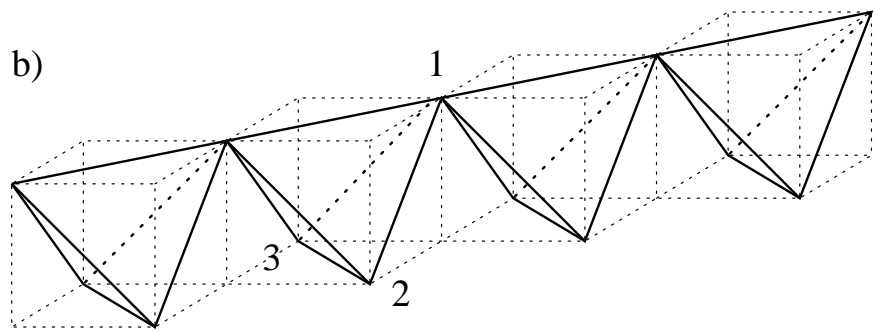
\begin{figure}
\caption{
Two illustrations of the Hubbard chain:
a) The filled circles denote the sites,
the solid lines indicate the bonds, where
hopping is possible. The dashed line frames
one unit cell. Sites within a unit cell
are labeled by an index $\alpha=1,2,3$.
b) The topology of the chain may also be
visualized by a chain of connected tetrahedra.
}
\label{fig:1}
\end{figure}

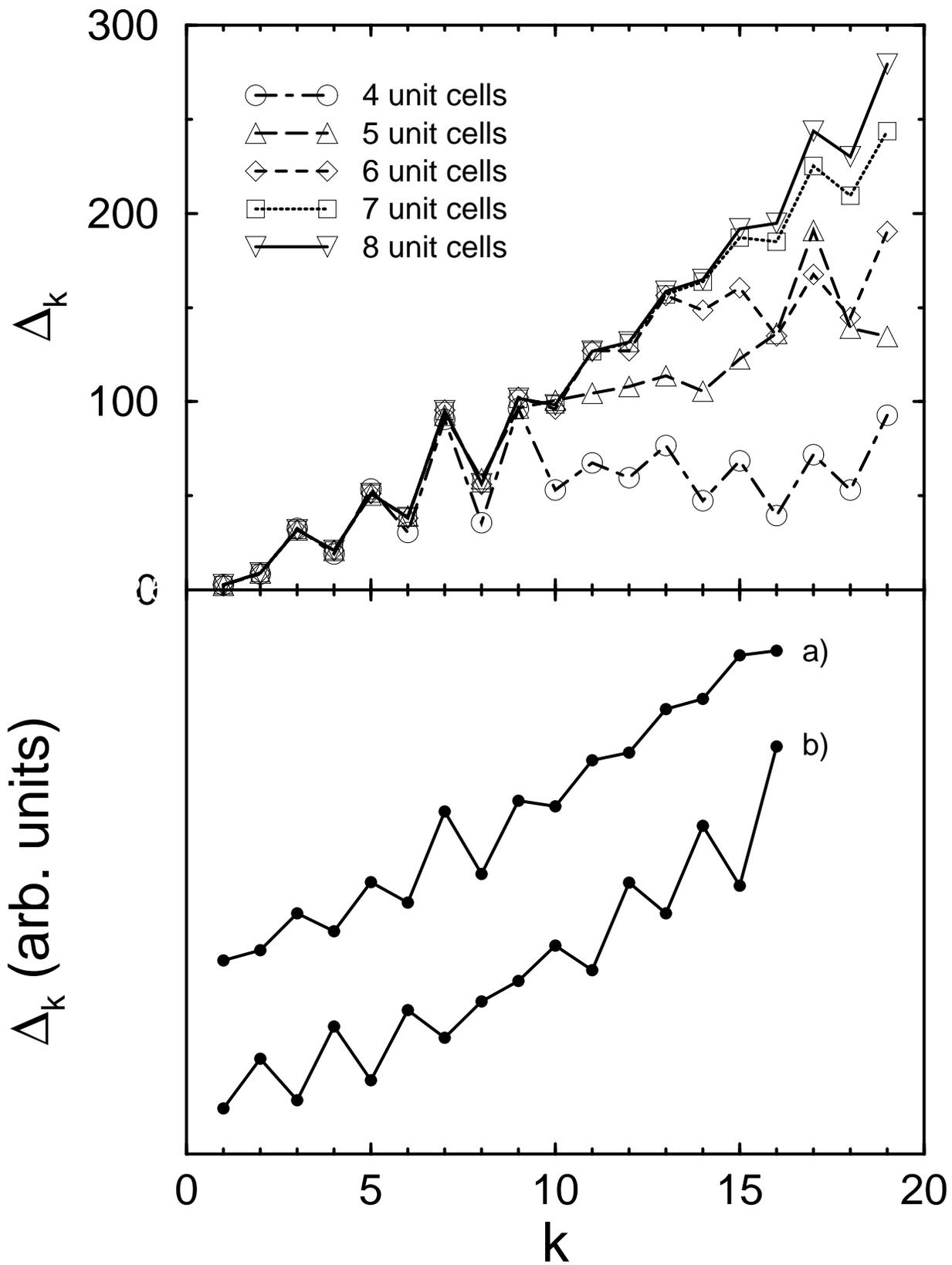
\begin{figure}
\caption{
Continued-fraction coefficients $\Delta_k$ verus $k$.
The upper graph shows coefficients obtained for the
$q=0$ spin degree of freedom ${\rm A}=\sum_l S^z_{l,\alpha=1}$
for various system sizes $N=4-8$ with particle numbers
of $2N$.
The lower graph shows two generic patterns
of coefficients. Data a) are identical to the spin data
for 8 unit cells of the upper graph. The set b)
displays results for the $q=0$ charge degree of freedom
${\rm A}=\sum_{l,\sigma} n^{}_{l,1,\sigma}$. Lines are
intended to guide the eye.
}
\label{fig:2}
\end{figure}

\begin{references}

\bibitem{hubbard}J.\,Hubbard,
                 Proc.\ R.\ Soc.\ London Ser. A
                 {\bf 276}, 238 (1963);
                 M.\,C.\,Gutzwiller,
                 Phys.\ Rev.\ Lett.
                 {\bf 10}, 159 (1963);
                 J.\,Kanamori,
                 Prog.\ Theor.\ Phys.
                 {\bf 30}, 275 (1963).

\bibitem{brandt-giesekus}U.\,Brandt and A.\,Giesekus,
                         Phys.\ Rev.\ Lett.
                         {\bf 68}, 2648 (1992).

\bibitem{strack}R.\,Strack,
                Phys.\ Rev.\ Lett.
                {\bf 70}, 833 (1993);
                R.\,Strack and D.\,Vollhardt
                Phys.\ Rev.\ Lett.
                {\bf 70}, 2637 (1993).

\bibitem{bares-lee}P.-A.\,Bares and P.\,A.\,Lee
                 Phys.\ Rev.\ B
                 {\bf 49}, 8882 (1994).

\bibitem{tasaki1}H.\,Tasaki,
                 Phys.\ Rev.\ Lett.
                 {\bf 70}, 3303 (1993).

\bibitem{tasaki2}H.\,Tasaki,
                 Phys.\ Rev.\ B.
                 {\bf 49}, 7763 (1994).

\bibitem{pauling}L.\,Pauling,
                 Proc.\ R.\ Soc.\ London A
                 {\bf 196}, 343 (1949);
                 P.\,W.\,Anderson,
                 Mater.\ Res.\ Bull.
                 {\bf 8}, 153 (1973);
                 P.\,Fazekas and P.\,W.\,Anderson,
                 Philos.\ Mag.
                 {\bf 30},423 (1974).

\bibitem{haydock}R.\,Haydock,
                 Solid State Phys.
                 {\bf 35}, 215 (1980).

\bibitem{lee}M.\,H.\,Lee,
             Phys.\ Rev.\ B
             {\bf 26}, 2547 (1982).

\bibitem{gagliano}E.\,R.\,Gagliano, E.\,Dagotto,
                  A.\,Moreo, and F.\,C.\,Alcaraz
                  Phys.\ Rev.\ B
                  {\bf 34}, 1677 (1986).

\bibitem{viswanath_jap}V.\,S.\,Viswanath and G.\,M\"uller,
                   J.\ Appl.\ Phys.
                   {\bf 67}, 5486 (1990).

\bibitem{pettifor}{\em The recursion method and its applications},
                  edited by D.\,G.\,Pettifor and D.\,L.\, Weaire
                  (Springer-Verlag, New York, 1985).

\bibitem{vs-gm}V.\,S.\,Viswanath and G.\,M\"uller,
               {\em The Recursion Method -- Application
               to Many-Body Dynamics},
               Springer-Verlag Berlin Heidelberg (1994).

\bibitem{lanczos}C.\,Lanczos,
                 J.\ Res.\ Nat.\ Bur.\ Stand.
                 {\bf 45}, 255 (1950).

\bibitem{viswanath}V.\,S.\,Viswanath, S.\,Zhang,
                   J.\,Stolze, and G.\,M\"uller
                   Phys.\ Rev.\ B
                   {\bf 49}, 9702 (1994).

\bibitem{hilbert}The corresponding integral [see Eq.\ (\ref{eq:17})]
                 leads to a sum of the complex error function and
                 the complex exponential integral.

\bibitem{bradbury}W.\,W.\,Bradbury and A.\,Fletcher,
                  Num.\ Math.
                  {\bf 9}, 259 (1966).

\bibitem{nightingale}M.\,P.\,Nightingale, V.\,S.\,Viswanath,
                     and G.\,M\"uller,
                     Phys.\ Rev.\ B
                     {\bf 48}, 7696 (1993).
\end{references}
\end{document}